# Nonuniform magnetic domain-wall synapses enabled by population coding

Ya Qiao, Yajun Zhang, and Zhe Yuan[*]

Center for Advanced Quantum Studies and Department of Physics, Beijing Normal University, Beijing 100875, China

Traditional artificial intelligence implemented in software is usually executed on accurate digital computers. Nevertheless, the nanoscale devices for the implementation of neuromorphic computing may not be ideally identical, and the performance is reduced by nonuniform devices. In biological brains, information is usually encoded by a cluster of neurons such that the variability of nerve cells does not influence the accuracy of human cognition and movement. Here, we introduce the population encoding strategy in neuromorphic computing and demonstrate that this strategy can overcome the problems caused by nonuniform devices. Using magnetic memristor device based on current-induced domain-wall motion as an example, we show that imperfect storage devices can be applied in a hardware network to perform principal component analysis (PCA), and the accuracy of unsupervised classification is comparable to that of conventional PCA using ideally accurate synaptic weights. Our results pave the way for hardware implementation of neuromorphic computing and lower the criteria for the uniformity of nanoscale devices.

---

[*] zyuan@bnu.edu.cn

# I. Introduction

Neuromorphic computing has attracted increasing attention in realizing efficient brain-like computing hardware that is stimulated by the successful applications of neural networks in artificial intelligence [1]. As a typical example, the crossbar-array-based structure is extensively studied with nonvolatile memory devices as synapses at the crossover positions [2-4]. By embedding the storage modules into the computing unit, crossbar structures exhibit the characteristics of in-memory computing to break the "von Neumann bottleneck". Currently, storage modules are usually implemented by memristors, i.e., electric devices with variable resistance, using resistive [5-7], phase change [8-10], and magnetic memories [11-13]. The resistance of a magnetic device is controlled by specific magnetic configurations, which can be manipulated by an external magnetic field and/or an electric current. Therefore, magnetic memories usually have better endurance than other types of memristors [14-16].

Magnetic tunnel junctions are technically mature magnetic devices for digital storage that exhibit low or high resistance depending on the relative alignment between two magnetic layers [17, 18]. Such two states representing "0" and "1" are perfect for digital memory, but it is difficult to reach and maintain a magnetic state stabilized between low and high resistances. The synaptic weight needs to be finely tuned in supervised or unsupervised learning [19, 20], and this requirement is fulfilled by magnetic textures with continuously variable magnetization. For example, both magnetic domain walls [21] and skyrmions [22-25] have been proposed as promising candidates for magnetic synapses in neuromorphic computing. The domain walls can be shifted back and forth to realize multi-level resistance as the synaptic weight, while the number of skyrmions can be artificially modulated to vary the resistance. In particular, magnetic skyrmions have the advantages of small size, low driving current density. For the domain walls, there are many different types of devices, such as the domain-wall-based magnetic tunnel junctions [26] and the Hall bar devices [27,28] we will consider later. However, one of the common striking technical difficulties in the application of these magnetic analogue synapses is the nonuniformity in devices. Here we take the domain-wall-based Hall bar device as an example without loss of generality. When a magnetic domain wall in a CoFeB thin film with perpendicular magnetic anisotropy (PMA) is continuously moved by a longitudinal electrical current, the total magnetization is gradually varied, resulting in a continuously varying anomalous Hall resistance [27,28]. The latter can be experimentally measured using Hall bar devices. In experiment, a moving domain wall may be pinned somewhere in the sample by random local barriers due to, e.g., structural defects and inhomogeneous alloy components introduced in the fabrication [29,30]. Edge modulation and notches can improve the deterministic control [31-33]. Nevertheless, the randomness still exists and the resulting local barriers of notches are not ideally identical.

In biological neurons, synapses are not precisely the same, and the action potentials passed through synapses may contain random noise. Even if the presynapse does not receive a stimulation, the postsynapse may still have a spontaneous weak potential response caused by random events in the synaptic transmission mechanism [34], such as spontaneous release of intracellular $Ca^{2+}$ ions and noise in ion channels, spontaneous vesicle release and fusion with membranes. Synaptic noise and variability induce uncertainty in the subsequent neuronal firing pulses. Nevertheless, information precision in brains is hardly affected by this synaptic variability during neural dynamics, mainly because of the brain's strategy of population coding. Experimentally, population coding has been observed in visual and motor cells [35, 36]. The information is effectively duplicated via, e.g., processing simultaneous transmission from multiple neurons [37] or to next neurons via multiple axons [38]. This strategy indicates that information is not encoded by individual cells but by a group of cells; therefore, it is very robust against variability or even damage to single cells [39].

Instead of simply duplicating the nerve pathway, it is essentially much better to employ bell-shaped tuning curves with different preferences distributed in the receptive field to increase coding efficiency [40]. In practice, a group of biased superparamagnetic magnetic tunnel junctions have been used as the basic units of neuronal population coding because they have nonlinear tuning curves of flipping frequency as a function of the input electric voltage and/or current [41, 42]. The redundancy of synapses is associated with the redundancy of neurons using multiple storage devices to represent the synaptic weight from an individual input to a particular output [43, 44]. The purpose is to improve the tolerance for variability of storage devices. For linear classification, such redundancy effectively increases the dimension of input data to help the subsequent networks distinguish input information [45].

In this paper, we demonstrate that the population coding strategy allows nonuniform storage devices to be applied in neuromorphic computing with high precision. As an example, we take disordered magnetic domain walls to store the synaptic weights in a simple cross bar network to perform principal component analysis (PCA) [46]. Micromagnetic simulation shows that the direct application of nonuniform magnetic domain wall storage fails to classify the mouse protein dataset, while population coding is found to effectively increase the tolerance of nonuniform storage devices. Our findings provide a general solution in neuromorphic computing with variable hardware, which is applicable for magnetic devices as well as resistive and phase-change devices. The rest of this paper is organized as follows. We introduce the population coding strategy using magnetic devices by analogy to biological neural systems in Sec. II, where the current-driven dynamics of magnetic domain walls simulated using micromagnetics are presented in detail. In Sec. III, we took the mouse protein dataset and performed PCA using a standard cross bar structure. In Sec. IV, ideally precise synapses and disordered synapses made of magnetic domain walls are considered. Although classification cannot be achieved using conventional PCA with disordered magnetic synapses, population coding for the input data eventually enables the task with precision comparable to that of ideal synapses. A short summary is given in Sec. V.

## II. Brain-inspired population coding strategy

### A. General concept of population coding

Reliable information transfer in the brain depends on the complex and redundant connections of cells. Pulse signals pass through the synaptic structure composed of the axon of the previous cell and the dendrite of the next cell, while the cell body of the neuron integrates the received messages and emits pulses of action potential. Figure 1(a) shows that three input neuron cells are connected to one target cell via the synaptic structure, which is marked by the circle and magnified in the inset. With the same pulse stimulus received, the three neurons may generate slightly different impulse responses, which are then transferred to the postsynaptic neuron. The messages transferred through synapses are inevitably affected by different degrees of variability, which arise from, e.g., the random noise caused by spontaneous weak current in the synapse itself. However, such variability does not lower the precision of information transfer and processing in the brain, mainly because information is encoded in the brain by populations or clusters of cells. This encoding strategy is called population coding. Although each cell is affected by independent variability, the random noise can be averaged and eliminated at the end of the neural pathways [34].

Population coding expresses neural signals through joint activities of many neurons, where the neurons are not simply duplicated to have the same response to the stimuli. Instead, each neuron exhibits the strongest response to a certain input in the reception field, which is called its preferred stimulus. When the stimulus gradually moves away from the preferred stimulus, the response will decrease. Such a group of bell-shaped tuning curves effectively maximizes the Fisher information in informatics and hence maximizes the coding efficiency with limited number of neurons [47]. In computational neurosciences, the bell-shaped tuning curves of cellular responses can be approximated by a Gaussian function [36, 48]. We follow this approximation and use a set of Gaussian functions with the same variance as shown in Fig. 1(c) for population encoding, where the maxima of these Gaussians are uniformly distributed in the reception field. For an input stimulus $x$, the $n$ neurons show different response strengths $X_1(x), X_2(x), \cdots, X_n(x)$ to encode this external stimulus, as schematically illustrated by the vertical dashed line in Fig. 1(c). The neuronal response signals $X_1(x), X_2(x), \cdots, X_n(x)$ are then transferred to the next layer of neurons in a feedforward network via synaptic connections, as sketched in Fig. 1(b). Thus, the input signal $x$ is encoded by multiple neurons with the population coding strategy, where $n$ synaptic connections are needed.

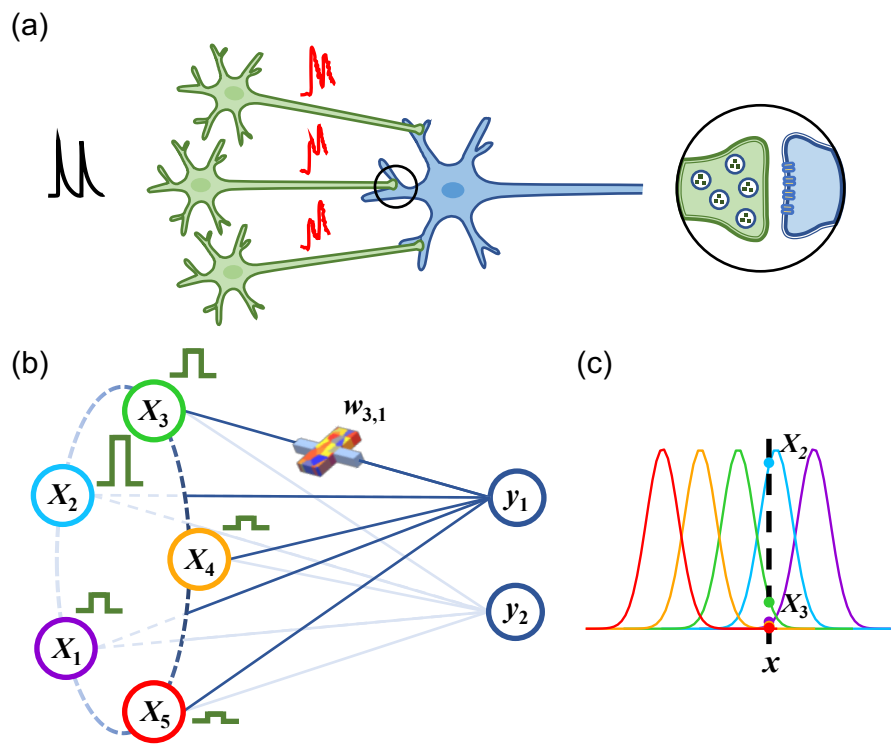

Fig. 1. (a) Schematic of signal transmission in biological neuron cells. Three input cells (green) and one target cell (blue) are connected through synapses, and the magnified plot of a synapse is shown in the inset with vesicles and receptors. Even with the same

stimulus received, the output signals transferred to the target cell can be different. (b) Schematic diagram of a feedforward network structure with the population coding strategy for the input signals. After receiving the same input stimulus $x$, the neurons in the first layer exhibit different responses $X_i$, which are transferred to the second layer via all-to-all connections. (c) Gaussian tuning curves of the group of neurons performing population encoding. The vertical dashed line illustrates the corresponding response $X_i$ of each neuron for the same input stimulus $x$.

### B. Storage devices with magnetic domain walls

To examine the advantage of the population coding strategy, we consider four-terminal synaptic storage devices containing a domain wall in a perpendicular ferromagnetic material [28, 49]. By shifting the domain-wall position via spin transfer torque (STT) [50] or spin orbit torque (SOT) [51, 52], we can in principle continuously vary the anomalous Hall resistance to mimic the synaptic weight. However, storage devices using magnetic domain walls are often criticized for their low reliability. The inevitable disorder in the material generates randomly distributed pinning sites for domain-wall motion, which results in irregular pinning and depinning processes [53, 54]. Although the SOT has the advantages of lower power consumption and faster velocity [55], we apply the STT to drive the domain wall for simplicity in micromagnetic simulation. The four-terminal storage device based on STT-driven domain wall motion and anomalous Hall detection is schematically plotted in Fig. 2(a). Here, we choose the ferromagnetic material CoFeB reported in the literature with perpendicular magnetic anisotropy [56]. The magnetization on both sides of the domain wall is either upward (along +z) or downward (along -z). As driven by the STT, the Bloch and Néel types of domain walls have approximately the same dynamical properties and the Bloch walls are employed in our simulation. For the domain wall located at the center of the device, the measured Hall signal vanishes. When an electric current is applied along the domain wall, the resulting STT drives the domain wall motion [50]. Then, the upward and downward magnetization in the devices is no longer balanced, which can be quantitatively measured from the Hall resistance/voltage [57].

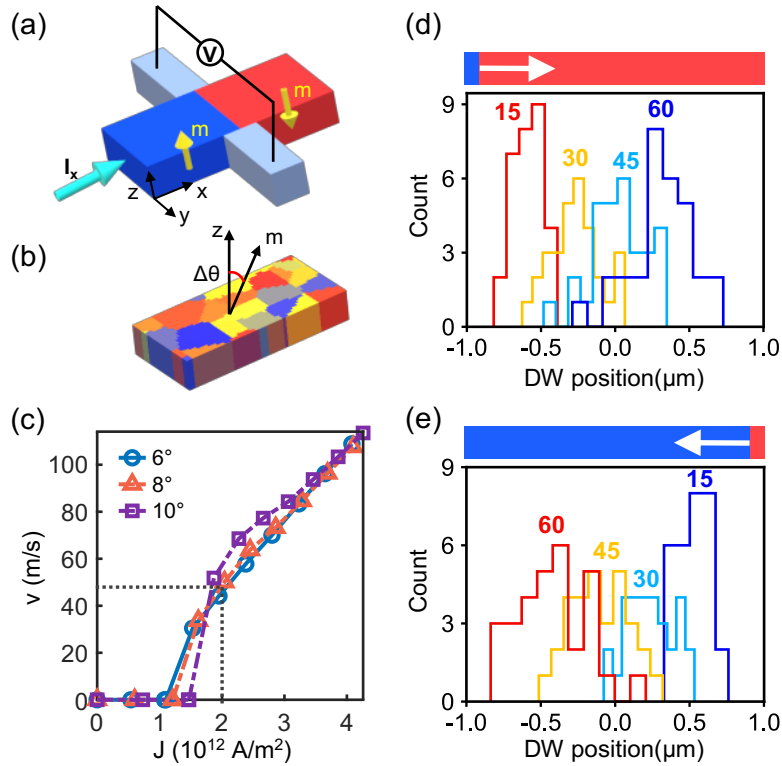

Fig. 2. (a) Schematic of the four-terminal Hall bar device based on magnetic domain walls. The current direction is applied in the longitudinal (*x*) direction, while the Hall voltage is measured in the transverse (*y*) direction. (b) Voronoi diagram of a magnetic material. The sample is divided into regions with an average diameter of approximately

10 nm. The easy axis of magnetic anisotropy in each region randomly deviates from the $z$-axis with the standard deviation $\Delta\theta$. (c) Calculated domain-wall velocity as a function of the driving electrical current density $J$ for different $\Delta\theta$. The black dotted lines show the applied current density for tuning the synaptic weights and the corresponding domain wall velocity. (d) and (e) Simulated domain-wall position distribution as a function of current pulse number starting from the left (d) and right boundary (e). The number of pulses is marked at the top of each distribution curve.

The measured Hall resistance in the four-terminal device is essentially proportional to the magnetization component along the z-axis, and a continuously varying resistance can be obtained by moving the domain wall via the current-induced torque. Simulating current-driven domain walls is achieved by solving the Landau-Lifshitz-Gilbert equation, including the STT terms [58, 59],

$$\dot{\boldsymbol{m}} = -\gamma \boldsymbol{m} \times \boldsymbol{H}_{\text{eff}} + \alpha \boldsymbol{m} \times \dot{\boldsymbol{m}} - (\boldsymbol{u} \cdot \nabla)\boldsymbol{m} + \beta[\boldsymbol{m} \times (\boldsymbol{u} \cdot \nabla)\boldsymbol{m}] \qquad (2)$$

where $\gamma$ is the gyromagnetic ratio, $\alpha$ is the Gilbert damping constant and $\boldsymbol{H}_{\text{eff}}$ is the effective magnetic field consisting of the exchange, anisotropy and external fields. $\boldsymbol{u} = PJ\mu_B/eM_s$ represents the magnitude of the STT, where $P$ is the spin polarization, $J$ is the current density, and $\beta$ is a nonadiabaticity parameter. Here, we employ the micromagnetics software package MUMAX3 [60] to simulate the domain wall motion driven by STTs, where the CoFeB sample is 2 μm long, 80 nm wide and 6 nm thick. This thickness is much larger than the usual values for the experimental samples of CoFeB with the perpendicular magnetic anisotropy [49]. It is chosen to gain a relatively fast computation and good numerical convergence and would not change any conclusions drawn in this study. We choose the exchange interaction constant $A=3\times10^{-11}$ J/m, the perpendicular magnetic anisotropy $K=8\times10^5$ J/m$^3$ and the saturation magnetization $M_s=8\times10^5$ A/m [56]. The anomalous Hall resistivity is 3.22 μΩ cm at 300 K [27] resulting in the maximum Hall resistance of 215 mΩ. In practice, the domain wall has the boundary to keep the perpendicular magnetization within the range of [-0.9, 0.9]$M_s$ and the resulting Hall resistance is in the range of [-195, 195] mΩ. To include the random disorder in ferromagnetic materials, we take the Voronoi diagram method in our micromagnetic simulation [54]. As schematically illustrated in Fig. 2(b), the ferromagnetic metal is divided into polygonal regions with an average diameter of 10 nm. Each Voronoi cell represents a crystal grain, and its easy axis slightly deviates from the ideal perpendicular anisotropy. Statistically, these deviations for all Voronoi cells follow the normal distribution with its standard deviation $\Delta\theta$.

In real experimental samples, ferromagnetic materials always contain defects, impurities and other types of disorder, and the domain wall can thus be pinned at some positions unless the electrical current density is larger than the critical value to overcome the pinning potential. The domain wall velocity is plotted in Fig. 2(c) as a function of the density of the driving electric current, where the circles, triangles and squares correspond to the standard deviation in the easy axis tilting $\Delta\theta=6°$, 8° and 10°, respectively. With increasing disorder, the calculated critical current density increases monotonically from $1.10\times10^{12}$ to $1.48\times10^{12}$ A/m$^2$, which is comparable with the reported experimental values from $0.32\times10^{12}$ to $1\times10^{12}$ A/m$^2$ in literature [50, 61-63].

With small disorder, the calculated domain wall velocity above the critical current density shows a linear dependence on the current density [58, 59]. However, this linear dependence does not hold with relatively strong disorder, and the nonlinearity is significant for $\Delta\theta =8°$ and 10°, as shown in Fig. 2(c). Again, the nonlinear current-density-dependence indicates a much stronger disorder in our simulation compared to those in experimental samples. In experiments, there is still a relatively linear relationship between the velocity and the current density [64, 65]. It is worth noting that the current density and velocity range we plotted in Fig. 2(c) is far below the Walker breakdown, which corresponding to a large velocity of approximately 660 m/s using the parameters in our simulation [58].

We deliberately choose the worse cases than the recent experimental situation to examine the capability of the proposed population coding strategy. In the following, we take $\Delta\theta =8°$ in the simulations of storage devices with magnetic domain walls in our networks. A synapse requires that the weight stored in the device can be continuously varied during the training process. This continuous tunability is tested by injecting current pulses with a fixed density of $2\times10^{12}$ A/m$^2$ and a time interval of 0.5 ns. The pulses gradually push the domain wall, resulting in a continuous variation in the Hall resistance. Here, we consider 30 disordered domain-wall devices, in which the initial position of the domain wall is set at the left boundary corresponding to the Hall resistance of -195 mΩ. After injecting certain number of electrical current pulses, the domain walls are moved towards the right end and the distribution of the final position are plotted in Fig. 2(d). With increasing the number of pulses from 15 to 60, the distribution of the domain wall position gradually moves to the right side. Then we repeat this process with the initial position of the domain wall at the right boundary corresponding to the Hall resistance of 195 mΩ. The distribution of the final position is shown as a function of the number of opposite current pulses in Fig. 2(e), which gradually moves towards the left end with increasing the pulse number. These devices exhibit significant diversity in the number of current pulses for one cycle, and these diverse devices will be used below in the simulation of a hardware neural network.

### III. PCA network with nonuniform storage devices

PCA is a statistical method for cluster analysis for a dataset consisting of a large number of interrelated variables [66]. The key idea of PCA is to look for the uncorrelated and ordered principal components (PCs), which are linear combinations of original variables but retain most of the variation present in all original variables. By directly calculating the eigenvector of the covariance matrix, the first PC is obtained in the direction of the maximum variance. Then the second PC can be found in the direction, which is orthogonal to the first PC and has the largest variance. Thus, using the first few PCs, one can significantly reduce the dimensionality of the dataset and achieve the best clustering in the low-dimensional space.

PCA can be realized using a cross bar as a simple network [46], which has high-dimensional input $x_i$ and reduced output $y_j$. Here, we only consider the first two PCs and therefore take just two output neurons. As shown in Fig. 3, there is a four-terminal domain-wall-based storage synapse at every crossing, and the synaptic weight is updated during the training process. When an electric voltage is applied in the direction of the horizontal bars, the current is injected into the domain wall from the left side because the other side is kept grounded. A voltmeter measures the Hall voltage perpendicular to the magnetic domain wall, which accumulates over the horizontal bars as the output signals.

We use a set of Gaussian functions with the fixed variance of 0.18, whose mean values are uniformly distributed from $-n$ to zero. The reception field covers the whole range of the input variables $x$. As schematically illustrated in the left inset of Fig. 3, when an input variable $x = -1.3$ is encoded using the population coding strategy by $n$ neurons, $x$ is converted into $X_1 = 0.01, X_2 = 0.73, X_3 = 0.25, \cdots$, respectively. Then the converted variables $X_1, X_2, \cdots, X_n$ are in turn transformed to the electrical voltage $V_1, V_2, \cdots, V_n$ applied to the horizontal bars. Here the amplitudes of voltage are proportional to $X_i$ but is always kept much smaller than the threshold voltage to prevent the domain wall motion. Then, the current density flows through the domain wall, and the Hall voltage can be read by the voltmeter. The accumulated Hall voltages in the two vertical output terminals are the first and second PCs, respectively. If the population coding strategy is not applied, the input variables are directly transformed to the voltages applied on the horizontal bars. Then, the crossbar network performs the standard PCA.

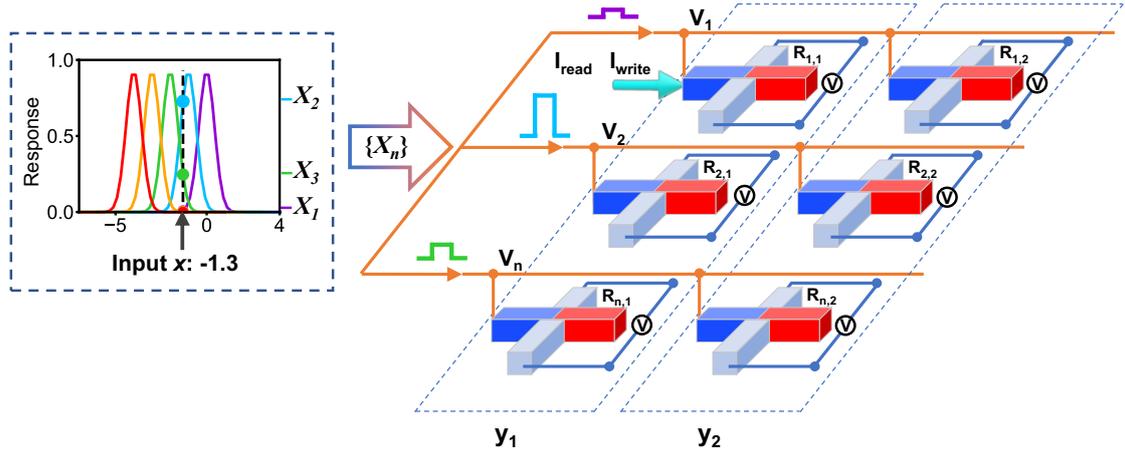

Fig. 3. Crossbar network to perform PCA with storage devices based on magnetic domain walls. Left inset: The input variable $x$ is encoded by $n$ neurons using the population coding strategy. The resulting group of variables $X_1, X_2, \dots, X_n$ are converted into voltages $V_1, V_2, \cdots, V_n$ applied on different horizontal bars. The voltage accumulation of each device in the two vertical columns is the first and second principal components, respectively. In the training process, the synaptic weight is updated by injecting the required writing current to adjust the domain wall position.

The synaptic weights $w_{ij}$ in this network are defined in the range from -1 to 1, which are mapped to the Hall resistances [-195, 195] mΩ of the domain walls. During the operation of the PCA network, the encoded voltages $V_1, V_2, \cdots, V_n$ are too small to move the magnetic domain wall, such that the synaptic weights are kept. Then, the signals from the two output terminals are read out.

The training of this PCA network is an unsupervised learning process that follows Sanger's rule [67], which is generalized Hebbian learning. The weight $w_{ij}$ is updated after every input and output by changing a small amount

$$\Delta w_{ij} = \alpha y_j\left(x_i - \sum_{k=1}^{j} w_{ik} y_k\right),$$

where α=0.005 represents the learning rate, and the cumulative outputs $y_j = \sum_{i=1}^{2} x_i w_{ij}$ correspond to the two PCs. The synaptic weights can be regarded as two vectors $w_{i1}$ and $w_{i2}$, which should be mutually orthogonal and have lengths of unity after convergence [67, 68]. In the training process, we choose a current density that is $2.0 \times 10^{12}$ A/m² higher than the critical current with a pulse duration of 0.5 ns. Hall resistance variation can be achieved by injecting a certain number of current pulses $N_{\text{write}}$ to push the domain wall to a desired position. For every device with a magnetic domain wall, the variation $\Delta w_{ij} = \frac{\Delta R_H}{195 \text{ m}\Omega} = \frac{\partial R_H}{\partial N} \frac{N_{\text{write}}}{195 \text{ m}\Omega}$ with the slope $\frac{\partial R_H}{\partial N}$ determined by fitting the linear relationship between the Hall resistance and pulse number for each device. However, the synaptic weights usually cannot be ideally tuned by the required number of pulses $N_{\text{write}}$ to the expected value due to the disordered domain walls, as seen in Fig. 2(d) and (e). The uncertainties will be examined in the next section.

**IV. Simulation results and Discussion**

In this section, we consider three types of PCA networks: (1) a conventional PCA with ideally accurate synaptic weights, (2) a network that uses storage devices based on disordered domain walls, and (3) a network using domain-wall-based storage combined with the population coding strategy. We selected the mouse protein expression level dataset [69] to test the performance. This dataset documents the measured protein expression levels of 10 mice after memantine injection. There are 300 samples in the dataset, half of which are used for training and the other half for testing. Every sample has 4 protein expression values as the input parameters. Some mice were stimulated to learn (context-shock) and others were not, and the output of PCA is to classify these two groups.

Using the conventional PCA method with ideally accurate synaptic weights, we apply the corresponding voltages converted from the four input parameters of each sample to the horizontal bars. Then, the $4 \times 2$ synapses are updated according to Sanger's rule until the weights converge. The norm of the ideal weight vectors for the first ($|w_{i1}|$) and second PCs ($|w_{i2}|$) are plotted in Fig. 4(a) as a function of the training step. Both norms gradually approach unity, and their mutual angle converges to 90°, as shown in the inset. In the inference stage, the 150 testing samples are analyzed by the PCA network, and the two PCs of each sample are plotted in Fig. 4(c). Then, we apply logistic regression to find the dashed line, which classifies the two different groups with an accuracy of 90%.

Then, we replace the ideally accurate synaptic weights with four-terminal storage devices based on disordered magnetic domain walls and repeat the PCA network training. The two vector norms using domain-wall-based storage are plotted in Fig. 4(b) as a function of the training step. While $|w_{i1}|$ converges to unity after 200 training steps, $|w_{i2}|$ nevertheless tends to saturate at approximately 0.7, far from unity due to the large uncertainties of the domain-wall-based storage. The problem is also reflected in the mutual angle of the two vectors, which fluctuates at approximately 53°. After the training, we obtain the two PCs shown in Fig. 4(c), and the clustering accuracy by the same logistic regression is 78%, significantly lower than the conventional PCA. Therefore, the performance of PCA clustering is reduced by nonuniform devices based on the disordered domain walls.

Generally, the performance of PCA may depend on the initial weights; thus, we randomly generate 500 groups of initial weights for both ideally accurate weights or those based on domain-wall storage. Figure 4(d) shows the average accuracy of clustering as a function of the training step. The conventional PCA with ideally accurate weights gradually improves the clustering accuracy during training. The average accuracy converges to more than 95%, and its standard deviation decreases. However, for the synaptic weights stored in disordered domain walls, the average accuracy is hardly improved by training, which is significantly lower than the conventional PCA. The large and nearly unchanged standard deviation suggests that the initial weights have a large influence.

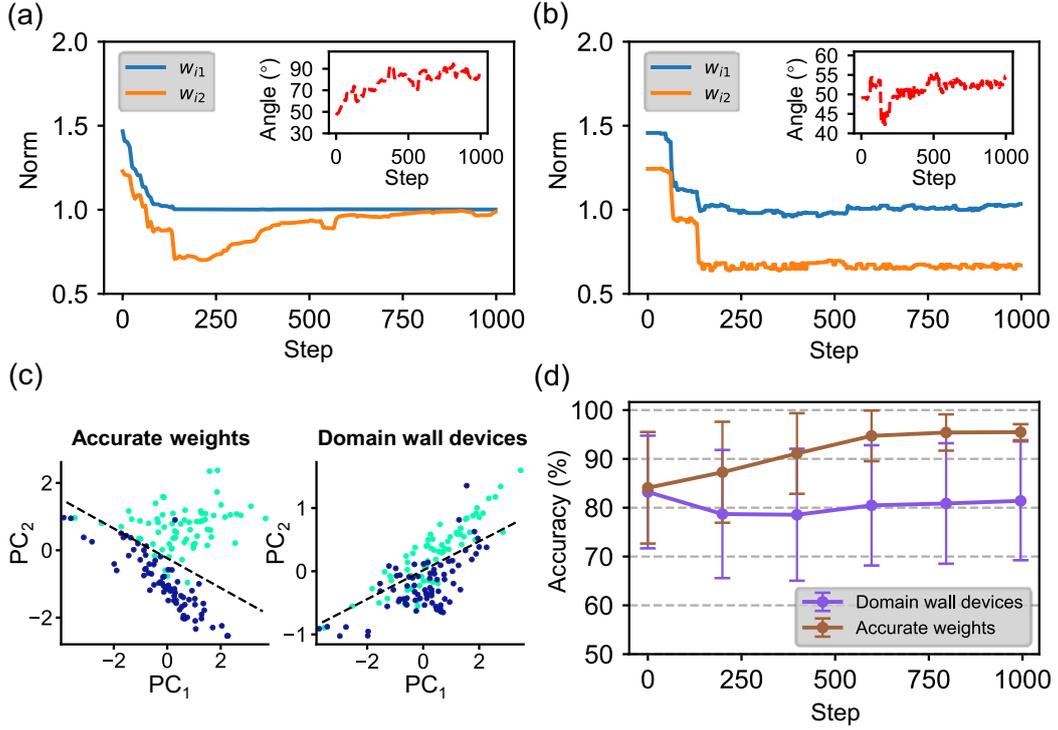

Fig. 4. The norm of vectors for synaptic weights $|w_{i1}|$ and $|w_{i2}|$ as a function of training step using ideally accurate synapses (a) and nonuniform storage devices based on magnetic domain walls (b). The blue and orange solid lines show the weight vector norms corresponding to the first and second PCs, respectively. The insets of (a) and (b) show the mutual angle of the two weight vectors. (c) Calculated PCs using the trained PCA network for the 150 testing samples, with accurate weights and nonuniform storage based on disordered domain walls, respectively. The dashed lines are the classification boundary obtained by the logistic regression. (d) Clustering accuracy of PCA for 500 different groups of initial weights.

Now, we apply the population coding for the PCA network with nonuniform storage devices based on magnetic domain walls, as shown in Fig. 3. The input data are coded using $n$ neurons, whose preferences are uniformly distributed in the whole reception field. The resultant groups of input voltages $V_1, V_2, \cdots, V_n$ are applied to the horizontal bars of the network, and the weights are updated using Sanger's rule in the training process. After training, the 150 testing samples of the mouse protein expression dataset are used to check the clustering accuracy. To exclude the influence of initial weights, we also repeat the training and inference check with 500 different groups of initial weights. Figure 5(a) shows the average clustering accuracy using a different number of neurons (*n*=4, 8 and 20) for the population coding. At *n*=4, the clustering accuracy is already improved compared with the purple line in Fig. 4(d). The average accuracy increases to nearly 90% with n=8 and approximately 95% with *n*=20. The accuracy exhibits strong fluctuation with large *n* because more disordered storage devices are involved, but the accuracy is noticeably improved. We summarize the average accuracy and its standard deviation as a function of the number of neurons used in population coding in Fig. 5(b). The average clustering accuracy increases monotonically with an increasing number of neurons, and the standard deviation gradually shrinks, indicating the convergence of population coding with respect to the number *n*. At *n*=20, the performance of the PCA network with nonuniform storage devices is comparable to that of conventional PCA using ideally accurate storage. Considering the fact that the dispersion of devices used in our numerical simulation is even worse than that of typical devices in experiments, the population coding strategy is demonstrated to overcome the problem caused by nonuniform devices in neuromorphic hardware implementation.

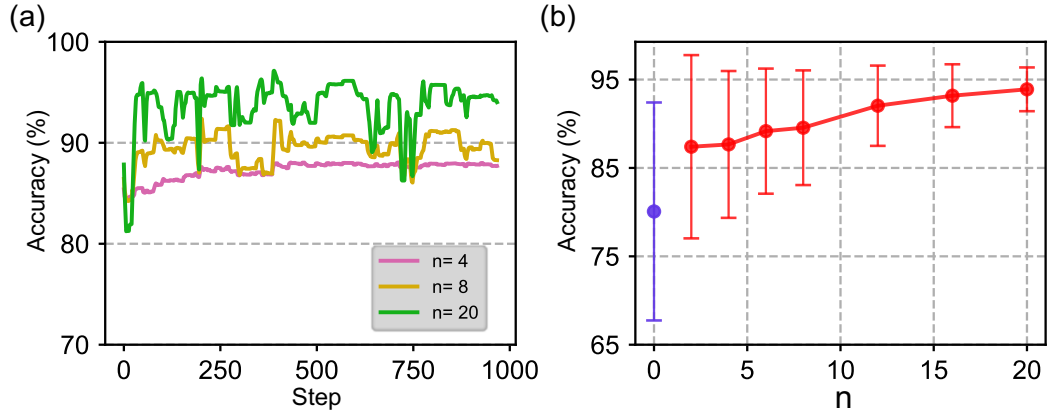

Fig. 5. (a) Clustering accuracy of the PCA network with domain-wall-based storage using population coding. The magenta, yellow and green lines correspond to the population coding using 4, 8, and 20 neurons, respectively. (b) Clustering accuracy as a function of the number of neurons in the population coding strategy. The purple dot represents the clustering result obtained using disordered domain wall storage without population coding.

## V. Conclusions

To summarize, we have introduced a biomimetic population coding scheme in neuromorphic computing for eliminating the influence of nonuniform hardware devices. Ferromagnetic domain walls have been proposed as the magnetic counterpart of memristors, while the disorder in ferromagnetic materials reduces the accuracy in controlling the position of domain-wall motion, resulting in nonuniformity in data storage. Taking the four-terminal magnetic memristor as an example, which is tuned by current-induced STT and is read out by measuring the Hall resistance, we first demonstrate that the simple PCA network cannot work properly using these nonuniform devices.

In biological brains, some information is encoded with a group of neurons using the strategy of population coding. We apply this population coding to input information into the PCA network with nonuniform storage devices. To maximize the coding efficiency, the preferences of these neurons are uniformly distributed in the whole reception field. By increasing the number of neurons in population coding, the PCA network with highly nonuniform storage devices can achieve the accuracy of unsupervised classification by the conventional PCA method with ideally accurate data storage. Our findings in this work shed light on how to improve the robustness of neuromorphic computing, especially for hardware devices that are slightly unidentical due to the difficulty in fabrication and micromachining technology. Although we have employed the domain-wall-based Hall bar devices in this work, the conclusions are applicable for other synaptic devices using magnetic domain walls or magnetic skyrmions.

## Acknowledgements

This work was supported by National Natural Science Foundation of China (Grants No. 11734004 and No. 12174028).